\documentclass[preprint,authoryear,11pt]{elsarticle}

\usepackage{graphicx,subfigure}
\usepackage{epstopdf}
\usepackage{amsmath, amssymb}
\usepackage{rotating}
\usepackage{morefloats}
\usepackage{comment}
\usepackage{natbib}
\usepackage{caption}
\usepackage{booktabs}
\usepackage{setspace}
\usepackage[margin=2.5cm]{geometry}
\begin{document}
\title{Diversified reward-risk parity in portfolio construction}
\author[gs]{Jaehyung Choi\corref{cor}\fnref{disc1}}
\ead{jj.jaehyung.choi@gmail.com}

\author[citi]{Hyangju Kim\fnref{disc2}}
\ead{hyangju.kim@gmail.com}

\author[sbbus]{Young Shin Kim}
\ead{aaron.kim@stonybrook.edu}

\address[gs]{Goldman Sachs \& Co., New York, USA}
\address[citi]{Citigroup, Inc., New York, USA}
\address[sbbus]{College of Business, Stony Brook University, New York, USA}

\cortext[cor]{Corresponding author}
\fntext[disc1]{Disclosure: The opinions and statements expressed in this paper are those of the author and may be different to views or opinions otherwise held or expressed by or within Goldman Sachs. The content of this paper is for information purposes only and is not investment advice or advice of any other kind. None of the author, Goldman Sachs, or its affiliates, officers, employees, or representatives accepts any liability whatsoever in connection with any of the content of this paper or for any action or inaction of any person taken in reliance upon such content or any part thereof.}
\fntext[disc2]{Disclosure: The views, information, or opinions expressed in this publication are solely those of the authors involved and do not represent those of Citigroup, its respective affiliates, or employees.}

\begin{abstract}
	We introduce diversified risk parity embedded with various reward-risk measures and more generic allocation rules for portfolio construction. We empirically test the proposed reward-risk parity strategies and compare their performance with an equally-weighted risk portfolio in various asset universes. The reward-risk parity strategies we tested exhibit consistent outperformance evidenced by higher average returns, Sharpe ratios, and Calmar ratios. The alternative allocations also reflect less downside risks in Value-at-Risk, conditional Value-at-Risk, and maximum drawdown. In addition to the enhanced performance and reward-risk profile, transaction costs can be reduced by lowering turnover rates. The diversified reward-risk parity allocations gain superior performance in the Carhart four-factor analysis.
\end{abstract}
\begin{keyword}
	portfolio construction, asset allocation, risk parity, reward-risk measures, classical tempered stable distribution, ARMA, GARCH
	
	\emph{JEL classification:} G11 \sep G12 \sep C58
\end{keyword}
\maketitle

\section{Introduction}
    Since Markowitz's seminal paper (\cite{Markowitz:1952}), portfolio construction has been one of the most intriguing topics in finance. Not confined to academia, it has been settled as a crucial investment procedure for practitioners who seek higher profitability and control financial risks under various investment constraints in prevailing market conditions. In particular, the Markowitz framework makes the portfolio construction process evolve into a more mathematical and quantitative decision-making process. Markowitz's portfolio construction process is basically solving an optimization problem to `optimally' allocate capital by maximizing or minimizing objective functions that reflect investors' investment goals and risk appetites. Finding extrema in the optimization problems leads the investors to the optimal solution for their investment goals. For example, the global minimum variance portfolio (\cite{Markowitz:1952}) tries to minimize a variance or to maximize an expected return at a given level of risk-return profiles. Similarly, \cite{Tobin:1958} and \cite{Sharpe:1964} also demonstrated that the highest Sharpe ratio is an optimal point on the efficient frontier in the presence of the risk-free asset.

    However, there are several drawbacks to the Markowitz portfolio. First, the Markowitz portfolios are sensitive to changes in model input parameters (\cite{Michaud:1989}). For example, even when the Markowitz model inputs -- a multivariate expected return vector or a covariance matrix of an investment universe -- are slightly changed, the resulting portfolio allocation to each constituent can vary abruptly. As the model parameters are updated with newly-arrived market data to portfolio optimization tools, the Markowitz framework potentially demands massive portfolio reallocation such that causes expensive transaction costs which erode portfolio performance. Second, the existence of optimal solutions under investment constraints is not always guaranteed. It is often hard to find the optimal solution satisfying a given constraint set in feasible domains. Although these disadvantages can be alleviated by the help of shrinkage estimation (\cite{Ledoit:2003}), regularization (\cite{DeMiguel:2009a}), and constraint impact (\cite{Jagannathan:2003}), the limitations could be still persistent.
	
	A more fundamental and philosophical question to the Markowitz framework is that the portfolio variance is not a representative of genuine risks to investment portfolios because larger positive returns contribute to increase profitability and volatility simultaneously. As potential solutions, other types of objective functions have been suggested for resolving the shortcomings and issues in the Markowitz-type optimization. Genuine portfolio risk measures have been exploited as objective functions in order to incorporate actual downside portfolio risks into asset allocation. For example, finding portfolio weights such that minimize Value-at-Risk (\cite{Gaivoronski:2005}), conditional Value-at-Risk (\cite{Rockafellar:2000,Krokhmal:2002}), and maximum drawdown instead of standard deviation is more compatible and consistent with investment insights on downside risks.
	
	Since the financial crisis in 2008, risk parity strategies have received increasing attention from both academia and industry. While the traditional `well-diversified' portfolios failed substantially in this market turmoil, the risk parity strategies relatively performed well. This is because the risk parity portfolio focuses on the allocation of risk in order to construct a truly diversified portfolio. For example, \cite{Bruder:2012} introduced the portfolio construction by matching the risk contribution of each asset to the same level. 
	
	However, it could be difficult for some practitioners to implement risk budgeting methodologies in real world because setting the equal risk budgeting based on a given risk measure would be computationally and practically demanding to achieve. For example, some risk parity methods require partial derivative of risk measures with respect to weight in order to calculate the risk contribution (\cite{Stoyanov:2006, Kim:2012}). Other risk parity portfolio construction methods such as maximum diversification ratio (\cite{Choueifaty:2008}) and maximum decorrelation (\cite{Christoffersen:2010}) also need robust matrix inversion and regularization techniques rooted from matrix analysis and statistics particularly when relevant covariance matrix and correlation matrix are high-dimensional. For larger asset universes, the situation becomes much more challenging due to the curse of dimensionality. Moreover, more complex models are more prone to sensitivity to input parameter changes. For example, it is difficult for several well-known portfolios to outperform the equally-weighted portfolio and even if those complex portfolios can outperform the naive allocation, such performance could be derived from estimation errors (\cite{DeMiguel:2009a, DeMiguel:2009b}).
	
	To mitigate these technical difficulties in risk parity methodologies, simple and heuristic approaches to risk parity also have been developed. Starting with the equally-weighted portfolio that is apparently the simplest portfolio, the market capitalization-weighted portfolio (\cite{Sharpe:1964}) is another example of the simple diversification methods. \cite{Fernholz:1982} also suggested another market capitalization-based allocation in which the weight of each asset is a power of its market capitalization. It is plausible to exploit other referential data such as fundamental weights based on accounting information (\cite{Arnott:2005}). Additionally, the equal risk contribution (ERC) strategies (\cite{Maillard:2010}) bet against market exposures via beta in the capital asset pricing model (CAPM). The diversified risk parity (DRP) portfolios suggested by \cite{Maillard:2010} adopt the standard deviation of each asset as the portfolio weights. Similar to the DRP portfolios, the diversified minimum variance (DMV) portfolio allocates weights inversely-proportional to variance. Still, these approaches are not based on reward-risk measures nor traditional risk measures of each individual asset.
	
    In this paper, we suggest heuristic allocation methodologies based on various reward-risk measures and general forms of allocation functions. The portfolio construction rules are empirically tested in multiple investment universes. These alternative portfolios as the generalization of the ERC allocation and the DRP allocation provide various advantages in performance and risk profile over the equally-weighted portfolio. Additionally, this paper is seeking alternative portfolios that are as simple and straightforward in construction as the equally-weighted allocation but outperform the naive allocation. However, this paper is not a simple blind search for outperforming portfolios based on a horse race-type comparison between various diversified reward-risk parity portfolios and the benchmark portfolio. Instead, this paper extends the scope of diversified risk parity to reward-risk measures and more generic allocation rules in order for investors to consider diverse options to construct their portfolios that are more in line with their investment goals and risk tolerances. 

    In the next section, alternative allocation rules using diversified reward-risk parity are introduced. In section \ref{drrp_reward_risk}, various reward-risk measures and risk models for the portfolio construction are provided. In section \ref{drrp_data_method}, data sets and methodologies for this study are given. In section \ref{drrp_result}, we present performance and reward-risk profiles of alternative allocations. In section \ref{drrp_factor_analysis}, the Carhart four-factor analysis on the portfolio performance is conducted. In section \ref{drrp_conclusion}, we conclude the paper.

\section{Diversified reward-risk risk party allocations}
\label{drrp_theory}
	In this section, we introduce the diversified reward-risk parity as the generalization of diversified risk parity by using reward-risk measures and various allocation rules.
	
	One simplest risk parity portfolio is the equal risk contribution (ERC) portfolios (\cite{Maillard:2010}). In an investment universe with $N$ assets, the allocation weight of each asset in the ERC portfolio is given by
    \begin{eqnarray}
    	\label{wgt_erc}
    	w_i=\frac{\beta_i^{-1}}{\sum_{i=1}^{N}\beta_i^{-1}}
    \end{eqnarray}
    where $\beta_i$ is the CAPM beta of the $i$-th asset. Considering the financial interpretation of the CAPM beta, the ERC portfolio penalizes exposures to the market portfolio. By betting against the beta, it tries to reduce asset-level correlations to the benchmark.

	\cite{Maillard:2010} also suggested the diversified risk parity (DRP) allocation based on standard deviation. The weights in the DRP portfolio are allocated with the inverse of volatility: 
    \begin{eqnarray}
    	\label{wgt_drp}
    	w_i=\frac{\sigma_i^{-1}}{\sum_{i=1}^{N}\sigma_i^{-1}}
    \end{eqnarray}
    where $\sigma_i$ is the standard deviation of the returns for the $i$-th asset. In this portfolio construction, less volatile investment vehicles are more preferred to more volatile ones. 
    
    Similarly, it is also possible to leverage variance instead of standard deviation. The diversified minimum variance (DMV) portfolio allocates more weights on assets with smaller variance: 
    \begin{eqnarray}
    	\label{wgt_dmv}
    	w_i=\frac{\sigma_i^{-2}}{\sum_{i=1}^{N}\sigma_i^{-2}}
    \end{eqnarray}
    where $\sigma_i$ is the standard deviation of the returns for the $i$-th asset. The DRP portfolio and the DMV portfolio penalize the volatility of each asset in the form of standard deviation and variance, respectively.

	In principle, the portfolio construction rules based on the heuristic risk parity such as Eq. (\ref{wgt_erc}), Eq. (\ref{wgt_drp}), and Eq. (\ref{wgt_dmv}) simply assign more weights on assets with smaller risks than assets with larger risks based on the assumption that CAPM beta and volatility are the proxies of risks. By favoring less risky assets over more risky assets, the allocation rules decide the extent of how much penalization is given to each asset based on risk measures. In these allocation schemes, the investors are risk-averse. 
	
	The various allocation weights mentioned above can be generalized into the following form:
	\begin{eqnarray}
    	\label{wgt_inverse}
    	w_i=\frac{\rho_i^{-1}}{\sum_{i=1}^{N}\rho_i^{-1}}
	\end{eqnarray}
	where $\rho$ is a reward-risk measure of interest. For example, the ERC portfolio of Eq. (\ref{wgt_erc}) corresponds to $\rho=\beta$. In case of the DRP allocation and the DMV allocation, the weights are decided by $\rho=\sigma$ and $\rho=\sigma^2$ in Eq. (\ref{wgt_drp}) and Eq. (\ref{wgt_dmv}), respectively.
	
	As a further generalization of Eq. (\ref{wgt_inverse}), we extend diversified reward-risk parity allocations by a using more generic allocation rule $\Phi$ such as
	\begin{eqnarray}
	\label{wgt_general}
	w_i=\frac{\Phi(\rho_i)}{\sum_{i=1}^{N}\Phi(\rho_i)}.
	\end{eqnarray}
	The previous examples are special cases of this generalization, Eq. (\ref{wgt_general}). For example, the ERC allocation and the DRP allocation rules reflect $\Phi(\rho) =\rho^{-1}$. The DMV portfolio construction is considered in two different ways: the inverse square function of $\Phi(\rho) =\rho^{-2}$ with $\rho=\sigma$ or the inverse function $\Phi(\rho) =\rho^{-1}$ with $\rho=\sigma^2$. It is also noteworthy that the equally-weighted portfolio is a special case of a constant $\Phi$. Additionally, it is also natural to consider reward-risk measures in similar diversified risk parity portfolio construction.
	 
	The allocation rule $\Phi$ in Eq. (\ref{wgt_general}) defines how assets are assigned based on reward-risk measures. In principle, $\Phi$ can be any functions to model investors' investment preference, risk aversion, and investment horizon. For example, some investors prefer assets with better reward-risk measures such as cumulative returns in the past because they are more trend-followers. Opposite to the momentum investors, another market participant penalizes the same assets with belief and prediction that such trends will be diminished and reverted. Combining these two approaches, the others would invest his/her capitals to the assets with large reward-risk measures until the criteria hit the threshold. After then, weights on the assets beyond (below) the limit start to be decreased (increased). Furthermore, different lengths of investment windows would demand different allocation rules $\Phi$ even for the same reward-risk measures (\cite{Choi:2014}). 
	
	In this paper, we confine $\Phi$ to monotonically decreasing or increasing functions of $\rho$. Additionally, let us assume that we are risk-aversion investors. First of all, these assumptions pursue the simplicity and consistency of implementation across various reward-risk measures. Instead of adopting a very bespoke $\Phi$, diversified reward-risk parity relies on more straightforward allocation rules. Moreover, the simpler allocation rule makes clearer and more robust explanations on the validity of the new diversification method than any bespoke allocation rules.
	
	With $\Phi$ as monotonically decreasing or increasing functions of $\rho$, it is possible to exploit various functions as long as the forms of the functions are consistent with characteristics of reward-risk measures. Decision between monotonically decreasing or increasing functions used for candidates to $\Phi$ needs further considerations on which measures are adopted as $\rho$. The nature and characteristics of $\rho$ in Eq. (\ref{wgt_general}) are the main factors to eligible forms of $\Phi$ that are consistent with the financial meaning and definition of each reward-risk measure. For example, when investors seek to minimize risks in the portfolio, risky assets are penalized in allocation. In order to reduce the exposures on such risky assets in new reward-risk parity methods, an expected allocation rule $\Phi$ needs to be monotonically decreasing functions of a risk measure $\rho$. Contrary to risk measures, when pure reward measures are employed in portfolio construction via Eq. (\ref{wgt_general}), investors prefer assets with larger reward-risk metrics and avoid assets with lower reward-risk measures. In this sense, $\Phi$ for reward measures should be a monotonically increasing function.
	
	Reward-risk measures are categorized into two classes in the following ways. The first class of reward-risk measures includes reward-risk ratio measures such as Sharpe ratio, Calmar ratio, STAR ratio, and Rachev ratio. Since higher ratio measures are more preferable and lower ratio measures are less preferable, $\Phi(\rho)$ should be a monotonically increasing function in $\rho$. 
	
	The simplest candidate for diversified reward-risk parity based on the reward-risk ratio measures is proportional to the measures given by
    \begin{eqnarray}
    	\label{wgt_reward_raw}
    	w_i=\frac{\rho_i}{\sum_{i=1}^{N}\rho_i}
    \end{eqnarray}
    where $\rho_i$ is a reward-risk measure of the $i$-th asset. Other variations such as an exponent of a reward-risk measure $\rho$, similar to \cite{Fernholz:1982}, are also available. 
    
    Another linear allocation rule is the linear version of Eq. (\ref{wgt_reward_raw}):
    \begin{eqnarray}
    	\label{wgt_reward_linear}
    	w_i=\frac{a+b\rho_i}{\sum_{i=1}^{N}(a+b\rho_i)}
    \end{eqnarray}
    where $a$ and $b$ are non-negative constants. It is straightforward to check that the equal weight allocation is a special case of this generalized reward-risk parity when $b=0$ in Eq. (\ref{wgt_reward_linear}). 
    
    Additionally, Eq. (\ref{wgt_reward_linear}) is decomposed into the equally-weighted allocation part and the deviation from the average reward-risk measure:
    \begin{eqnarray}
    	\label{wgt_reward_linear_decomp}
    	w_i=\frac{1}{N} + \frac{b(\rho_i-\bar{\rho})}{\sum_{i=1}^{N}(a+b\rho_i)}
    \end{eqnarray}
    where $a$ and $b$ are non-negative constants, and $\bar{\rho}$ is the average reward-risk measure. This expression shows that the linear allocation of a reward-risk measure fluctuates around the equally-weighted allocation.

	There is the possibility of reward-risk ratio measures to be negative values that cause short-selling in the portfolio. Some investors do not short-sell assets or are not allowed to be in short positions. Opposite to the Markowitz optimization, constraints on no short-selling in diversified reward-risk parity can be directly implemented in more straightforward ways by explicitly imposing $\rho_i|_{+}=max(\rho_i, 0)$ instead of using $\rho_i$ directly. For example, the introduction of the quasi-linear allocation rule to Eq. (\ref{wgt_reward_raw}) and Eq. (\ref{wgt_reward_linear}) represent the asset allocations with no short-selling constraints in the following forms:
    \begin{eqnarray}
    	\label{wgt_reward_linear_floor_all}
    	w_i=\frac{(a+b\rho_i)|_{+}}{\sum_{i=1}^{N}(a+b\rho_i)|_{+}}
    \end{eqnarray}
    or 
    \begin{eqnarray}
    	\label{wgt_reward_linear_floor}
    	w_i=\frac{a+b\rho_i|_{+}}{\sum_{i=1}^{N}(a+b\rho_i|_{+})}
    \end{eqnarray}
    where $a$ and $b$ are non-negative constants and $\rho_i|_{+}=max(\rho_i, 0)$. Each weight in Eq. (\ref{wgt_reward_linear_floor_all}) has a hard flooring at zero. Meanwhile, the minimal non-zero allocation is guaranteed in Eq. (\ref{wgt_reward_linear_floor}).

	Another class of reward-risk measures includes pure risk measures such as volatility, variance, maximum drawdown, VaR, and CVaR. Since these metrics represent volatility or losses, assets with larger measures are less favorable. Weights on assets with larger risk measures need to be smaller than those on assets with smaller risk measures. 
	
	There are the four simplest ways of achieving the penalization of risks in allocation. The first implementation is taking the inverse of risk measures as a weighting scheme. Given a risk measure, the diversified reward-risk parity of a pure risk measure $\rho$ is defined as
    \begin{eqnarray}
    	\label{wgt_risk_inverse}
    	w_i=\frac{\rho_i^{-1}}{\sum_{i=1}^{N}\rho_i^{-1}}
    \end{eqnarray}
    where $\rho_i$ is the risk measure of $i$-th asset. 
    
    Another way of penalizing risk measures in portfolio construction is as follows:
    \begin{eqnarray}
    	\label{wgt_risk_linear}
    	w_i=\frac{a-b\rho_i}{\sum_{i=1}^{N}(a-b\rho_i)}
    \end{eqnarray}
    where $a$ and $b$ are non-negative constants. If $a=1$, $b=1$, and the measure is one of VaR, CVaR, and maximum drawdown, $1-\rho$ can be considered the expected portfolio value after severe losses if there is no correlation between assets. 
    
    The third way of the risk penalization is to linearize Eq. (\ref{wgt_risk_inverse}) in the following form:
    \begin{eqnarray}
    	\label{wgt_risk_inverse_linear}
    	w_i=\frac{a+b\rho_i^{-1}}{\sum_{i=1}^{N}(a+b\rho_i^{-1})}
    \end{eqnarray}
    where $a$ and $b$ are non-negative constants. Similar to Eq. (\ref{wgt_reward_raw}) and Eq. (\ref{wgt_reward_linear}), Eq. (\ref{wgt_risk_linear}) and Eq. (\ref{wgt_risk_inverse_linear}) also cover the equally-weighted allocation as a special case of $b=0$. 
    
    Lastly, it is possible to consider the no short-selling constraint by using $\rho_i|_{+}=max(\rho_i, 0)$, which is the risk version of Eq. (\ref{wgt_reward_linear_floor_all}) and Eq. (\ref{wgt_reward_linear_floor}), although such cases for risk measures are relatively rare.
    
\section{Reward-risk measures and risk models}
\label{drrp_reward_risk}
In this section, we visit definitions of reward-risk measures that are building blocks for portfolio construction by diversified reward-risk parity. Additionally, we present risk models for calculating the reward-risk measures. 

\subsection{Reward-risk measures}
	Throughout this paper, various reward-risk measures are employed for portfolio construction, performance assessment, and risk management. We shortly cover definitions of these measures for further usages.

\subsubsection{Volatility and variance}
	Volatility assesses return fluctuations of a given financial time series, and we compute the volatility as either of two different definitions: the standard deviation from historical returns and the conditional volatility from ARMA-GARCH models. Additionally, we also consider unconditional/conditional variances from these two volatility measures.

\subsubsection{Sharpe ratio}
	Sharpe ratio (\cite{Sharpe:1964}) is one of the most well-known reward-risk ratios in finance and also broadly used by practitioners. Sharpe ratio is defined as the ratio of expected return to standard deviation of a return time series $r$:
	\begin{eqnarray}
	\label{eq_sharpe}
		\textrm{SR}=\frac{\mathbb{E}[r-r_f]}{\sigma[r-r_f]}
	\end{eqnarray}
	where $\mathbb{E}[\cdot]$ is the expectation value, $\sigma[\cdot]$ is the standard deviation, and $r_f$ is the risk-free rate. In the Markowitz framework, the tangency portfolio exhibits the highest Sharpe ratio among all feasible portfolios under the existence of the risk-free asset (\cite{Markowitz:1952,Sharpe:1964}). From investors' point of view, assets with higher Sharpe ratios are more preferred in portfolio construction.
	
	Similar to the usage of two definitions in volatility, two different Sharpe ratio definitions are used for portfolio construction and performance assessment. One is the traditional Sharpe ratio definition that is the ratio of empirical average return to historical standard deviation. This is mentioned as marginal/unconditional Sharpe ratio in the paper. Another is conditional Sharpe ratio calculated as the ratio of conditional mean to conditional volatility computed by ARMA-GARCH models.

\subsubsection{Maximum drawdown}
	Maximum drawdown (MDD) is the worst consecutive loss in a specified time period. The maximum drawdown of a portfolio is given by
	\begin{eqnarray}
	\label{eq_mdd}
		\mathrm{MDD}=-\overset{}{\underset{\tau\in(0,T)}{\textrm{min}}}\Big(\overset{}{\underset{t\in(0,\tau)}{\textrm{min}}} r(t,\tau)\Big)
	\end{eqnarray}
	where $r(t,t')$ is the return during the time period between $t$ and $t'$. It is the worst realized performance since the inception of a portfolio during a given investment horizon. Apparently, assets with smaller maximum drawdowns are favored by investors. Additionally, maximum drawdown also encodes the information on time evolution of a return time series opposite to other quantile-based risk measures. For example, even when we use a given return time series for risk calculation, different maximum drawdowns are obtained by different sequential orders of the time series.  

\subsubsection{Calmar ratio}
	Back to the definition of Sharpe ratio, Eq. (\ref{eq_sharpe}), the denominator of the ratio is the standard deviation which penalizes assets with larger volatility. However, losses are not the only source of volatility. Although positive returns are the origins of portfolio profitability, such returns also contribute to increasing volatility. Hence, the separation of volatility from downside risks is crucial in considering performance measures. 
	
	Calmar ratio (\cite{Young:1991}) is the ratio of return to maximum drawdown:
	\begin{eqnarray}
	\label{eq_calmar}
		\textrm{Calmar}=\frac{R}{\textrm{MDD}}
	\end{eqnarray}
	where $R$ is the realized cumulative return, and $\textrm{MDD}$ is the realized maximum drawdown of Eq. (\ref{eq_mdd}) within a given investment time horizon. Investors prefer assets with higher Calmar ratios to assets with lower Calmar ratios. 
	
	In the original definition of Calmar ratio (\cite{Young:1991}), the time window for return and maximum drawdown was given to 36 months. In this paper, the Calmar ratio is computed from the realized return and the maximum drawdown in the six-month window for portfolio construction. In order to assess portfolio performance and risk, the Calmar ratio presents the ratio of annualized return to maximum drawdown from the whole period since the inception of the portfolio.

\subsubsection{Value-at-Risk and conditional Value-at-Risk}
	Value-at-Risk (VaR) is the loss at a given quantile of a performance distribution. For a given return time series $r$, VaR of $(1-\eta)100\%$ ($0<\eta<1$) is defined as
	\begin{eqnarray}
	\label{eq_var}
		\textrm{VaR}\big((1-\eta)100\%\big)=-\textrm{inf}\{ l | P(r-r_f>l)\le1-\eta\}
	\end{eqnarray}
	where $r_f$ is the risk-free rate and $P$ is the cumulative distribution function of a given underlying probability distribution. VaR describes the quantile risk of a return distribution as a single number computed from historical data or simulated data. 
	
	Despite its simplicity, several shortcomings of VaR still exist. First of all, non-subadditivity for VaR implies that portfolio diversification could bring a worse VaR value than the weighted sum of VaRs from its components. It is obviously counter-intuitive to the investment insight that diversification can reduce investment risks.
	
	\cite{Rockafellar:2000, Rockafellar:2002} suggested conditional Value-at-Risk (CVaR), also known as average Value-at-Risk. CVaR is the average loss of extreme losses worse than a given quantile loss. By using the definition of VaR in Eq. (\ref{eq_var}), the CVaR at $(1-\eta)100\%$ is defined as
	\begin{eqnarray}
	\label{eq_cvar}
		\textrm{CVaR}\big((1-\eta)100\%\big)=\frac{1}{\eta}\int_0^\eta \textrm{VaR}\big((1-\zeta)100\%\big)d\zeta
	\end{eqnarray}
	where $0<\eta<1$. When a portfolio suffers from severe losses worse than a threshold, a CVaR value indicates the average of such extreme losses.
	
	One advantage of CVaR over VaR is its coherency (\cite{Artzner:1999}). Opposite to the characteristics of VaR, CVaR fulfills not only the subadditivity but also other properties of coherent risk measures. In addition to the coherency of CVaR, much information on the downside tail is incorporated into CVaR. For example, CVaR values from two portfolios would be different even if VaRs of those portfolios are the same. It is obvious that portfolios with thicker downside tails exhibit worse CVaR values than portfolios with thinner downside tails.

\subsubsection{Stable tail adjusted return ratio}
	As the similar intuition of Calmar ratio in Eq. (\ref{eq_calmar}), \cite{Martin:2003} suggested the stable tail adjusted return (STAR) ratio in order to enhance the concept of Sharpe ratio. Instead of adopting the standard deviation in Sharpe ratio, Eq. (\ref{eq_sharpe}), the ratio is penalized by CVaR that represents genuine downside risks. For a time series $r$, the definition of STAR ratio at $(1-\eta) 100\%$ ($0<\eta<1$)  is given by
	\begin{eqnarray}
	\label{eq_star}
		\textrm{STAR}\big((1-\eta) 100\%\big)=\frac{\mathbb{E}[r-r_f]}{\textrm{CVaR}\big((1-\eta) 100\%\big)}
	\end{eqnarray}
	where $r_f$ is the risk-free rate. It is straightforward that the standard deviation in the definition of Sharpe ratio is replaced with CVaR defined in Eq. (\ref{eq_cvar}). By using the STAR ratio, assets with less downside tail risks exhibit larger STAR ratios. Obviously, portfolios with higher ratios are more preferred.

\subsubsection{Rachev ratio}
	Although the STAR ratio in Eq. (\ref{eq_star}) removes the ambiguity between volatility and downside risk, the expected return in the nominator of the STAR ratio definition still includes the redundancy regarding the downside risks. Extreme downward events penalize the ratio not only by increasing the CVaR in the denominator but also by decreasing the expected return in the nominator. It is necessary to isolate downside risks from the reward measure.
	
	\cite{Biglova:2004} introduced Rachev ratio to segregate contributions of losses from the reward part. Instead of employing the expected return, Rachev ratio is defined as the ratio of expected upward tail gain to downward tail loss. It is represented with the ratio of two CVaRs from an original return time series $r$ used in the denominator and its sign-inverted return time series in the nominator:
	\begin{eqnarray}
	\label{eq_r}
		\textrm{Rachev}((1-\eta) 100\%,(1-\zeta) 100\%)=\frac{\textrm{CVaR}\big((1-\eta) 100\%\big) \textrm{ for } (r_f-r)}{\textrm{CVaR}\big((1-\zeta) 100\%\big) \textrm{ for } (r-r_f)}
	\end{eqnarray}
	where $0<\eta<1$, $0<\zeta<1$, and $r_f$ is the risk-free rate. It is obvious that an asset with a lower downside risk and a higher upside reward is more preferred.
	
\subsection{Risk models}
	In order to test the robustness of diversified reward-risk parity to risk model choice, we apply three different risk models for reward-risk calculation. The first risk model for reward-risk measures is the ARMA(1,1)-GARCH(1,1) model with classical tempered stable (CTS) innovations suggested by \cite{Kim:2010}, \cite{Kim:2010b}, and \cite{Kim:2011}. Since this model captures autocorrelation, volatility clustering, skewness, and heavy-tailedness of a financial time series, various applications of this model are found in portfolio management (\cite{Tsuchida:2012,Beck:2013,Georgiev:2015,Anand:2016}) and momentum strategy (\cite{Choi:2015}). 
	
	In $\mathrm{CTS}(\alpha, C_+, C_-,\lambda_+,\lambda_-,m)$ distributions, $m$ is the location parameter, $\alpha$ is the tail index, $C_+$ and $C_-$ are the scale parameters, $\lambda_+$ and $\lambda_-$ are the decay rates of the tails. The characteristic function of a CTS distribution is in the following form:
	\begin{align}
	\label{cts_charac_ftn}
		\phi(u)=&\exp\Big(ium-iu\Gamma(1-\alpha)(C_+\lambda_+^{1-\alpha}-C_-\lambda_-^{1-\alpha})\nonumber\\
		&+\Gamma(-\alpha)\big(C_+\big((\lambda_+-iu)^\alpha-\lambda_+^\alpha\big)+C_-\big((\lambda_-+iu)^\alpha-\lambda_-^\alpha\big)\big)\Big)
	\end{align}
	where $C_+, C_-, \lambda_+, \lambda_-$ are positive, $\alpha\in(0,2)$, $m\in\mathbb{R}$ and $\Gamma$ is the gamma function.
	
	As described in \cite{Kim:2011}, the parameters of the ARMA(1,1)-GARCH(1,1)-CTS model are estimated from the following steps. With an assumption that residuals of the ARMA(1,1)-GARCH(1,1) model are Student's \textit{t}-distributed, the ARMA-GARCH parameters are estimated from maximum likelihood estimation (MLE). The innovations obtained from the previous step are applied to CTS model parameter estimation by fast Fourier transformation and MLE (\cite{Rachev:2011}). Using all the estimated model parameters, we can calculate various reward-risk measures. For more details, check \cite{Kim:2010, Kim:2010b, Kim:2011} and references therein.
	
	Another risk model for reward-risk measure calculation is a simple variant of the first model by replacing the CTS distribution as the innovation model with the normal tempered stable (NTS) distribution suggested by \cite{Barndorff:2001a} and \cite{Barndorff:2001b}. In this case, the risk model is the ARMA(1,1)-GARCH(1,1)-NTS model which is used in the literature (\cite{Anand:2016, Anand:2017,Kurosaki:2019}). The last risk model to use is the multivariate extension of the innovation distribution in the second model. With the multivariate normal tempered stable (MNTS) innovations (\cite{Kim:2012,Kim:2015,Kim:2021,Kim:2022}), we utilize DCC GARCH model (\cite{Engle:2001}) instead of the GARCH model to capture time-varying covariance matrix. In this case, the risk model for reward-risk measure calculation is the ARMA(1,1)-DCC GARCH(1,1)-MNTS model.
	 
\section{Dataset and methodology}
\label{drrp_data_method}

In this section, we introduce investment universes for empirically testing diversified reward-risk parity portfolio construction. After then, the portfolio construction procedure for empirical tests is explained.

\subsection{Dataset}
	In order to test diversified reward-risk parity allocation across various asset classes, we cover Dow Jones Industrial Average components, global equity benchmarks, largest mutual funds and ETFs, and SPDR U.S. sector ETFs. Adjusted daily price data sets of these asset universes were downloaded from Yahoo Finance.

\subsubsection{U.S. equity markets: Dow Jones Industrial Average}
	The constituents of the Dow Jones Industrial Average universe cover the period from January 1st, 1999 and December 31st, 2020. The tickers with component changes are listed in Table \ref{tbl_ticker_list_us_dji_yf_component}.

\begin{table}[!ht]
\centering
\caption{Ticker information for Dow Jones Industrial Average}
\label{tbl_ticker_list_us_dji_yf_component}
\resizebox{.5\paperheight}{!}{
\begin{tabular}{llll}
\toprule
Ticker &                                        Name & Minimum Date & Maximum Date \\
\midrule
    AA &                           Alcoa Corporation &   1999-01-01 &   2013-09-23 \\
  AAPL &                                  Apple Inc. &   2015-03-19 &   2020-12-31 \\
   AIG &          American International Group, Inc. &   2004-04-08 &   2008-09-22 \\
  AMGN &                                  Amgen Inc. &   2020-08-31 &   2020-12-31 \\
   AXP &                    American Express Company &   1999-01-01 &   2020-12-31 \\
    BA &                          The Boeing Company &   1999-01-01 &   2020-12-31 \\
   BAC &                 Bank Of America Corporation &   2008-02-19 &   2013-09-23 \\
     C &                              Citigroup Inc. &   1999-01-01 &   2009-06-08 \\
   CAT &                            Caterpillar Inc. &   1999-01-01 &   2020-12-31 \\
   CRM &                        Salesforce.Com, Inc. &   2020-08-31 &   2020-12-31 \\
  CSCO &                         Cisco Systems, Inc. &   2009-06-08 &   2020-12-31 \\
    DD &                     Dupont De Nemours, Inc. &   1999-01-01 &   2019-04-02 \\
   DIS &                     The Walt Disney Company &   1999-01-01 &   2020-12-31 \\
   DOW &                                    Dow Inc. &   2019-04-02 &   2020-12-31 \\
    GE &                    General Electric Company &   1999-01-01 &   2018-06-26 \\
    GM &                      General Motors Company &   1999-01-01 &   2009-06-08 \\
    GS &               The Goldman Sachs Group, Inc. &   2013-09-23 &   2020-12-31 \\
    GT &          The Goodyear Tire \& Rubber Company &   1999-01-01 &   1999-11-01 \\
    HD &                        The Home Depot, Inc. &   1999-11-01 &   2020-12-31 \\
   HON &                Honeywell International Inc. &   1999-01-01 &   2008-02-19 \\
   HON &                Honeywell International Inc. &   2020-08-31 &   2020-12-31 \\
   HPQ &                                     Hp Inc. &   1999-01-01 &   2013-09-23 \\
   IBM & International Business Machines Corporation &   1999-01-01 &   2020-12-31 \\
  INTC &                           Intel Corporation &   1999-11-01 &   2020-12-31 \\
    IP &                 International Paper Company &   1999-01-01 &   2004-04-08 \\
   JNJ &                           Johnson \& Johnson &   1999-01-01 &   2020-12-31 \\
   JPM &                        Jpmorgan Chase \& Co. &   1999-01-01 &   2020-12-31 \\
    KO &                       The Coca-Cola Company &   1999-01-01 &   2020-12-31 \\
  KRFT &                            Kraft Foods Inc. &   2008-09-22 &   2012-09-24 \\
   MCD &                      Mcdonald'S Corporation &   1999-01-01 &   2020-12-31 \\
   MMM &                                  3M Company &   1999-01-01 &   2020-12-31 \\
    MO &                          Altria Group, Inc. &   1999-01-01 &   2008-02-19 \\
   MRK &                           Merck \& Co., Inc. &   1999-01-01 &   2020-12-31 \\
  MSFT &                       Microsoft Corporation &   1999-11-01 &   2020-12-31 \\
   NKE &                                  Nike, Inc. &   2013-09-23 &   2020-12-31 \\
   PFE &                                 Pfizer Inc. &   2004-04-08 &   2020-08-31 \\
    PG &                The Procter \& Gamble Company &   1999-01-01 &   2020-12-31 \\
   RTX &           Raytheon Technologies Corporation &   2020-04-06 &   2020-08-31 \\
     T &                                   At\&T Inc. &   1999-11-01 &   2004-04-08 \\
     T &                                   At\&T Inc. &   2005-11-21 &   2015-03-19 \\
   TRV &               The Travelers Companies, Inc. &   2009-06-08 &   2020-12-31 \\
   UNH &             Unitedhealth Group Incorporated &   2012-09-24 &   2020-12-31 \\
   UTX &             United Technologies Corporation &   1999-01-01 &   2020-04-06 \\
     V &                                   Visa Inc. &   2013-09-23 &   2020-12-31 \\
    VZ &                 Verizon Communications Inc. &   2004-04-08 &   2020-12-31 \\
   WBA &              Walgreens Boots Alliance, Inc. &   2018-06-26 &   2020-12-31 \\
   WMT &                                Walmart Inc. &   1999-01-01 &   2020-12-31 \\
   XOM &                     Exxon Mobil Corporation &   1999-01-01 &   2020-08-31 \\
\bottomrule
\end{tabular}
}\caption*{The historical components of Dow Jones Industrial Average are listed with their tickers, names of companies, inclusion dates, and exclusion dates.}
\end{table}

\subsubsection{Other markets}
	For global equity benchmarks, daily price data in the period between January 1st, 1999 and December 31st, 2020 were downloaded. For the market universe of mutual funds and ETFs universe, we started from the list of largest mutual funds and ETFs starts from the top 25 largest mutual funds and ETFs list at Market Watch\footnote{https://www.marketwatch.com/tools/mutual-fund/top25largest}. Excluding money market funds and government securities funds, there are thirteen mutual funds and four ETFs between January 1st, 1999 and December 31st, 2020. For testing our methodologies in industry sectors, nine SPDR U.S. sector ETF tickers in the universe, which have existed since the inception of the SPDR sector ETF series, are chosen during the period between January 1st, 1999 and December 31th, 2020. 

\subsubsection{Risk-free rates}
	In order to calculate reward-risk measures introduced in the previous section, the yield of the 91-day U.S. Treasury bill is employed as the risk-free rate for the market universes mentioned above. In Yahoo Finance, the ticker of the 91-day U.S. Treasury bill is \textasciicircum IRX. The period of the daily data is from January 1st, 1999 to December 31st, 2020.
	
\subsection{Portfolio construction process}
	On a portfolio construction date, daily price data during the last six months are used for computing reward-risk measures. As mentioned in the previous section, these measures, except for maximum drawdown and Calmar ratio, are calculated by using three different risk models: the ARMA(1,1)-GARCH(1,1) models with CTS and NTS innovations, and the ARMA(1,1)-DCC GARCH(1,1)-MNTS model. Maximum drawdown and Calmar ratio are directly computed from historical return data.
	
	With reward-risk metrics from the past six months of the portfolio construction date, portfolio weights are decided by each allocation rule based on diversified reward-risk parity. In order to fairly compare performance with the equally-weighted portfolio, the short-selling caused by negative metrics is not allowed. 
	
	The allocation rules used for portfolio construction are given in Table \ref{tbl_allocation_rule}. First of all, the equally-weighted portfolio is the benchmark portfolio. For reward-risk ratio measures such as Calmar ratio, marginal/conditional Sharpe ratio, STAR ratio, and Rachev ratio, two allocation rules are used: $\rho_i|_{+}$ and $1+\rho_i|_{+}$. The first rule is flooring at zero that considers no short-selling and also corresponds to the case of $a=0, b=1$ in Eq. (\ref{wgt_reward_linear_floor}). The second rule is the linear version of the first rule that is the case of $a=1, b=1$ in Eq. (\ref{wgt_reward_linear_floor}). The constant term allows investors to impose positive weights across all the assets in the universe., i.e., at least non-zero weights to all the assets in the universe opposite to Eq. (\ref{wgt_reward_linear_floor_all}).
	
\begin{table}[!h]
\centering
\caption{Allocation rules by reward-risk and risk measures}
\label{tbl_allocation_rule}
\begin{tabular}{ll}
\toprule
Measure & Rule  \\
\midrule
	Equal Weigtht & 1 \\
  	Reward-risk ratio &   $\rho_i|_{+}$ \\
  	 &   $1+\rho_i|_{+}$ \\
   	Risk &   $1/\rho_i$ \\
  	 &   $1-\rho_i$ \\
\bottomrule
\end{tabular}
\end{table}	
	
	For pure risk measures such as maximum drawdown, unconditional/conditional volatility/variance, VaR, and CVaR, the allocation rules of $1/\rho_i$ and $1-\rho_i$ are exploited for penalizing downside risks during the portfolio construction. The former allocation is the case of $a=0$, $b=1$ in Eq. (\ref{wgt_risk_inverse_linear}) and the latter is $a=1$, $b=1$ in Eq. (\ref{wgt_risk_linear}). The flooring to avoid short-selling is not considered for risk measures because negative risk measures are usually not common.
	
	The alternative portfolios constructed by the diversification rules of reward-risk measures are being held for the next six months. After six months of holding, the portfolios are rebalanced based on the portfolio weights calculated by the metrics at that moment. The portfolio construction process occurs every month, and one-sixth of the overall portfolio is replaced by the processes described in the previous paragraphs.

\section{Results}
\label{drrp_result}

In this section, we provide performance results and risk profiles of various diversified reward-risk parity portfolios. In order to avoid repetitive description on the similar patterns found in different markets, we focus on the test results from the Dow Jones Industrial Average. Similarly, we present the results by the reward-risk measures from the ARMA(1,1)-GARCH(1,1) model with the CTS distribution. After then, we summarize common features and characteristics found in the diversified reward-risk parity portfolios across the various market universes and the other two risk models. 

\subsection{U.S. equity markets: Dow Jones Industrial Average}
	According to Table \ref{tbl_summary_stats_ratio_us_dji_yf}, it is noteworthy that diversified reward-risk parity based on ratio measures achieve enhanced performances than the equally-weighted portfolio of 9.84\% per year in U.S. Dow Jones Industrial Average universe. In particular, the Calmar ratio strategies are the most lucrative ones. The raw and linear allocation rules of Calmar ratio earn annually 10.32\% and 10.06\% on average, respectively. Additionally, the performance of both allocations is less volatile with standard deviations of 18.68\% and 19.49\% than the benchmark of 19.82\%. Moreover, the return distributions are more desirable with more negative skewness and smaller kurtosis than that of the benchmark.
	
	\begin{sidewaystable}[!ht]
\centering
\caption{Summary statistics for ratio portfolios in Dow Jones Industrial Average}
\label{tbl_summary_stats_ratio_us_dji_yf}
\resizebox{\textwidth}{!}{
\begin{tabular}{llrrrrrrrrrrrrrr}
\toprule
         Rule &      Measure &    Mean &     Std &    Skew &   Kurt. &   Cumul. &  Sharpe &  CSharpe &  Calmar &  Max DD &  VaR(95\%) &  CVaR(95\%) &  VaR95/99 &  CVaR95/99 &  Turnover \\[-2ex]\\
\midrule
            1 & Equal Weight &  9.8419 & 19.8158 & -0.0612 & 10.0569 & 442.6942 &  0.4967 &   3.0121 &  0.1702 & 57.8115 &     0.9079 &      1.2647 &    0.6120 &     0.6942 &    6.7482 \\[-2ex]\\
  $\rho|_{+}$ &       Sharpe &  9.5866 & 18.6657 & -0.1058 &  9.5262 & 438.7364 &  0.5136 &   2.5313 &  0.2178 & 44.0164 &     1.1694 &      1.6180 &    0.6175 &     0.6984 &   58.4629 \\[-2ex]\\
  $\rho|_{+}$ & Cond. Sharpe & 10.0084 & 18.9900 &  0.0777 &  9.4165 & 482.4326 &  0.5270 &   2.8412 &  0.2146 & 46.6281 &     1.0165 &      1.4015 &    0.6206 &     0.7011 &   66.8401 \\[-2ex]\\
  $\rho|_{+}$ &       Calmar & 10.3212 & 18.6839 & -0.0926 &  8.9297 & 530.3241 &  0.5524 &   2.4968 &  0.2616 & 39.4614 &     1.4439 &      1.9913 &    0.6202 &     0.7007 &   66.0731 \\[-2ex]\\
  $\rho|_{+}$ &   STAR(90\%) & 10.0513 & 18.9987 &  0.0734 &  9.2650 & 487.5955 &  0.5291 &   3.0306 &  0.2171 & 46.3046 &     0.9850 &      1.3596 &    0.6197 &     0.7006 &   69.2023 \\[-2ex]\\
  $\rho|_{+}$ &   STAR(95\%) &  9.9439 & 19.0295 &  0.0709 &  9.2443 & 473.4976 &  0.5226 &   3.0290 &  0.2147 & 46.3236 &     0.9835 &      1.3592 &    0.6186 &     0.6998 &   68.9955 \\[-2ex]\\
  $\rho|_{+}$ &   STAR(99\%) &  9.8527 & 19.0448 &  0.0688 &  9.2903 & 461.9973 &  0.5173 &   3.0236 &  0.2126 & 46.3333 &     0.9803 &      1.3551 &    0.6184 &     0.6995 &   68.5758 \\[-2ex]\\
  $\rho|_{+}$ & R(50\%,90\%) &  9.8592 & 19.5309 & -0.0313 &  9.8770 & 451.3793 &  0.5048 &   2.9391 &  0.1780 & 55.3774 &     0.9333 &      1.2985 &    0.6129 &     0.6948 &   18.7071 \\[-2ex]\\
  $\rho|_{+}$ & R(50\%,95\%) &  9.8020 & 19.5537 & -0.0335 &  9.8904 & 444.1117 &  0.5013 &   2.9308 &  0.1771 & 55.3396 &     0.9340 &      1.2994 &    0.6129 &     0.6949 &   18.2920 \\[-2ex]\\
  $\rho|_{+}$ & R(50\%,99\%) &  9.7292 & 19.5769 & -0.0358 &  9.9513 & 435.1502 &  0.4970 &   2.8570 &  0.1764 & 55.1599 &     0.9427 &      1.3107 &    0.6135 &     0.6953 &   18.8973 \\[-2ex]\\
  $\rho|_{+}$ & R(90\%,90\%) &  9.8749 & 19.6521 & -0.0504 &  9.8852 & 450.3813 &  0.5025 &   2.9520 &  0.1749 & 56.4495 &     0.9247 &      1.2871 &    0.6126 &     0.6946 &   13.7517 \\[-2ex]\\
  $\rho|_{+}$ & R(95\%,95\%) &  9.8542 & 19.6910 & -0.0559 &  9.8861 & 447.0405 &  0.5004 &   2.9514 &  0.1739 & 56.6507 &     0.9225 &      1.2838 &    0.6127 &     0.6947 &   13.0349 \\[-2ex]\\
  $\rho|_{+}$ & R(99\%,99\%) &  9.8703 & 19.7406 & -0.0621 &  9.8815 & 447.7668 &  0.5000 &   2.9025 &  0.1734 & 56.9068 &     0.9227 &      1.2836 &    0.6130 &     0.6949 &   14.0656 \\[-2ex]\\
$1+\rho|_{+}$ &       Sharpe &  9.8625 & 19.7864 & -0.0631 & 10.0528 & 445.7884 &  0.4984 &   3.0481 &  0.1711 & 57.6373 &     0.9037 &      1.2599 &    0.6112 &     0.6937 &    5.7905 \\[-2ex]\\
$1+\rho|_{+}$ & Cond. Sharpe &  9.8409 & 19.7445 & -0.0554 & 10.0114 & 444.2366 &  0.4984 &   3.0487 &  0.1717 & 57.3114 &     0.9021 &      1.2568 &    0.6118 &     0.6941 &    8.5247 \\[-2ex]\\
$1+\rho|_{+}$ &       Calmar & 10.0568 & 19.4868 & -0.0712 &  9.6921 & 476.2547 &  0.5161 &   3.2301 &  0.1829 & 54.9985 &     0.8789 &      1.2275 &    0.6096 &     0.6932 &   26.3331 \\[-2ex]\\
$1+\rho|_{+}$ &   STAR(90\%) &  9.8494 & 19.7620 & -0.0578 & 10.0215 & 444.8199 &  0.4984 &   3.0450 &  0.1714 & 57.4759 &     0.9026 &      1.2574 &    0.6119 &     0.6942 &    8.0382 \\[-2ex]\\
$1+\rho|_{+}$ &   STAR(95\%) &  9.8330 & 19.7745 & -0.0587 & 10.0247 & 442.6113 &  0.4973 &   3.0526 &  0.1709 & 57.5398 &     0.9019 &      1.2564 &    0.6119 &     0.6943 &    7.5883 \\[-2ex]\\
$1+\rho|_{+}$ &   STAR(99\%) &  9.8340 & 19.7865 & -0.0596 & 10.0370 & 442.4429 &  0.4970 &   3.0299 &  0.1707 & 57.6201 &     0.9056 &      1.2615 &    0.6119 &     0.6942 &    6.9639 \\[-2ex]\\
$1+\rho|_{+}$ & R(50\%,90\%) &  9.8461 & 19.7214 & -0.0530 &  9.9986 & 445.3800 &  0.4993 &   2.9885 &  0.1724 & 57.0988 &     0.9150 &      1.2741 &    0.6123 &     0.6944 &    8.7535 \\[-2ex]\\
$1+\rho|_{+}$ & R(50\%,95\%) &  9.8242 & 19.7396 & -0.0548 & 10.0095 & 442.4157 &  0.4977 &   3.0018 &  0.1718 & 57.1821 &     0.9130 &      1.2713 &    0.6123 &     0.6945 &    8.1285 \\[-2ex]\\
$1+\rho|_{+}$ & R(50\%,99\%) &  9.8101 & 19.7602 & -0.0569 & 10.0308 & 440.2824 &  0.4965 &   2.9766 &  0.1712 & 57.2862 &     0.9151 &      1.2743 &    0.6122 &     0.6944 &    7.4632 \\[-2ex]\\
$1+\rho|_{+}$ & R(90\%,90\%) &  9.8565 & 19.7297 & -0.0562 &  9.9690 & 446.3864 &  0.4996 &   2.9872 &  0.1725 & 57.1419 &     0.9156 &      1.2751 &    0.6122 &     0.6943 &    9.1059 \\[-2ex]\\
$1+\rho|_{+}$ & R(95\%,95\%) &  9.8456 & 19.7498 & -0.0589 &  9.9695 & 444.6470 &  0.4985 &   2.9762 &  0.1720 & 57.2379 &     0.9156 &      1.2750 &    0.6122 &     0.6944 &    8.7983 \\[-2ex]\\
$1+\rho|_{+}$ & R(99\%,99\%) &  9.8555 & 19.7742 & -0.0620 &  9.9653 & 445.2435 &  0.4984 &   2.9548 &  0.1718 & 57.3568 &     0.9155 &      1.2744 &    0.6125 &     0.6946 &    9.3972 \\[-2ex]\\
\bottomrule
\end{tabular}
}\caption*{Summary statistics for 6/6 monthly diversified reward-risk portfolios in Dow Jones Industrial Average are given in the table.  Mean and standard deviation are annualized numbers. Cumulative return is in percentage scale. Reward-risk and risk measures for 6/6 monthly diversified reward-risk portfolios in Dow Jones Industrial Average are also given. Sharpe ratio and conditional Sharpe ratio are annualized and Maximum drawdown (MDD) is in percentage scale. VaR and CVaR are represented in daily percentage scale. VaR 95/99 and CVaR 95/99 are from the ratio of 95\% to 99\%. The turnover rate is in percentage scale.}
\end{sidewaystable}

	Similar to the Calmar ratio portfolios, diversified Sharpe ratio allocations are also useful to construct portfolios that outperform the equally-weighted portfolio in average return and volatility. For example, the diversified conditional Sharpe ratio parity achieves annualized 10.01\% on average, the second strongest portfolio in profitability among all other portfolios including the benchmark. Additionally, the outperformance is based on smaller return fluctuations in the range of 18.67--19.74\%. Furthermore, the return distributions of the diversified Sharpe ratio strategies are thinner-tailed with smaller kurtosis indicating that the portfolios are less exposed to the probability of extreme losses.
	
	Diversified reward-risk parity strategies in STAR ratio and Rachev ratio are also more profitable with reduced volatility levels than the benchmark. Similar to the previous cases, the raw diversification rules of the reward-risk ratio measures tend to outperform the linear allocations of the same measures. For example, the raw STAR ratio (90\%)-based parity is the best STAR ratio portfolio with generating annually 10.05\% on average. Moreover, comparing with the benchmark volatility of 19.82\%, the standard deviation of the STAR ratio (90\%) allocation is decreased to 19.00\%. The similar pattern is also observed for all other raw and linear allocations of STAR ratio and Rachev ratio. The profitability of all the other diversified STAR (Rachev) ratio strategies is stronger in the range of 9.83--9.94\% (9.80--9.87\%) with lower volatility levels of 19.03--19.76\% (19.53--19.77\%). All the diversified STAR ratio and Rachev ratio portfolios are less exposed to extreme events with lower kurtosis values than the benchmark is.
	
	In Table \ref{tbl_summary_stats_risk_us_dji_yf}, diversified reward-risk parity allocations by pure risk measures such as VaR, CVaR, standard deviation, variance, and maximum drawdown earn comparable or weaker profitability than the equally-weighted allocation. The performance of linearly-diversified risk parity strategies is generally more profitable than inversely-diversified risk metric strategies. Meanwhile, the inverse risk metric portfolios are less volatile than the benchmark and the linear risk parity portfolios. For example, the performance of the linearly-diversified VaR and CVaR strategies is in the comparable range of 9.79--9.83\% while the volatility levels are decreased to 19.65--19.75\%. Although profits of the inverse VaR and CVaR portfolios decrease, volatility levels of these portfolios are also reduced. The similar pattern is found in the performance of the diversified standard deviation and variance allocations. However, there is notable reduction in annualized performance for the maximum drawdown allocations.  
	
\begin{sidewaystable}[!ht]
\centering
\caption{Summary statistics for risk portfolios in Dow Jones Industrial Average}
\label{tbl_summary_stats_risk_us_dji_yf}
\resizebox{\textwidth}{!}{
\begin{tabular}{llrrrrrrrrrrrrrr}
\toprule
    Rule &        Measure &   Mean &     Std &    Skew &   Kurt. &   Cumul. &  Sharpe &  CSharpe &  Calmar &  Max DD &  VaR(95\%) &  CVaR(95\%) &  VaR95/99 &  CVaR95/99 &  Turnover \\[-2ex]\\
\midrule
       1 &   Equal Weight & 9.8419 & 19.8158 & -0.0612 & 10.0569 & 442.6942 &  0.4967 &   3.0121 &  0.1702 & 57.8115 &     0.9079 &      1.2647 &    0.6120 &     0.6942 &    6.7482 \\[-2ex]\\
$1/\rho$ &   Max Drawdown & 9.4613 & 18.4023 & -0.0469 & 10.6753 & 430.0339 &  0.5141 &   3.5396 &  0.1878 & 50.3783 &     0.7981 &      1.1160 &    0.6088 &     0.6925 &   12.2917 \\[-2ex]\\
$1/\rho$ &           Std. & 9.5437 & 18.3332 & -0.0245 & 10.7354 & 441.0092 &  0.5206 &   3.0292 &  0.1870 & 51.0456 &     0.8418 &      1.1737 &    0.6114 &     0.6927 &    8.9994 \\[-2ex]\\
$1/\rho$ &       Variance & 9.4198 & 17.4372 & -0.0064 & 11.3465 & 445.2943 &  0.5402 &   3.2444 &  0.2030 & 46.4064 &     0.7900 &      1.1036 &    0.6100 &     0.6910 &   12.8717 \\[-2ex]\\
$1/\rho$ &     Cond. Std. & 9.4399 & 18.2613 & -0.0358 & 10.7230 & 430.5641 &  0.5169 &   3.2093 &  0.1869 & 50.5150 &     0.8147 &      1.1365 &    0.6108 &     0.6933 &   13.4904 \\[-2ex]\\
$1/\rho$ & Cond. Variance & 9.3563 & 17.4882 & -0.0352 & 11.4752 & 436.8391 &  0.5350 &   3.5107 &  0.2031 & 46.0674 &     0.7654 &      1.0719 &    0.6077 &     0.6904 &   18.7332 \\[-2ex]\\
$1/\rho$ &      VaR(90\%) & 9.5110 & 18.1583 & -0.0247 & 10.5880 & 440.9307 &  0.5238 &   3.2120 &  0.1908 & 49.8469 &     0.8213 &      1.1463 &    0.6104 &     0.6927 &   15.6314 \\[-2ex]\\
$1/\rho$ &      VaR(95\%) & 9.5415 & 18.1811 & -0.0288 & 10.6298 & 443.9930 &  0.5248 &   3.1987 &  0.1910 & 49.9530 &     0.8196 &      1.1435 &    0.6107 &     0.6929 &   14.2009 \\[-2ex]\\
$1/\rho$ &      VaR(99\%) & 9.4654 & 18.2269 & -0.0310 & 10.6545 & 434.2170 &  0.5193 &   3.1218 &  0.1898 & 49.8591 &     0.8238 &      1.1490 &    0.6110 &     0.6929 &   14.8284 \\[-2ex]\\
$1/\rho$ &     CVaR(90\%) & 9.4971 & 18.1974 & -0.0287 & 10.6387 & 438.4918 &  0.5219 &   3.1654 &  0.1904 & 49.8862 &     0.8220 &      1.1468 &    0.6108 &     0.6929 &   14.3708 \\[-2ex]\\
$1/\rho$ &     CVaR(95\%) & 9.4641 & 18.2186 & -0.0301 & 10.6451 & 434.2231 &  0.5195 &   3.1600 &  0.1898 & 49.8731 &     0.8209 &      1.1449 &    0.6111 &     0.6930 &   14.8309 \\[-2ex]\\
$1/\rho$ &     CVaR(99\%) & 9.4278 & 18.2491 & -0.0301 & 10.6794 & 429.4567 &  0.5166 &   3.0752 &  0.1894 & 49.7705 &     0.8271 &      1.1527 &    0.6117 &     0.6934 &   17.1826 \\[-2ex]\\
$1-\rho$ &   Max Drawdown & 9.4455 & 19.1010 & -0.0686 & 10.3003 & 413.5284 &  0.4945 &   3.1826 &  0.1754 & 53.8515 &     0.8530 &      1.1892 &    0.6113 &     0.6939 &    5.4928 \\[-2ex]\\
$1-\rho$ &           Std. & 9.8257 & 19.7671 & -0.0616 & 10.0639 & 441.9357 &  0.4971 &   3.0129 &  0.1707 & 57.5739 &     0.9062 &      1.2621 &    0.6120 &     0.6943 &    6.6901 \\[-2ex]\\
$1-\rho$ &       Variance & 9.8401 & 19.8121 & -0.0613 & 10.0568 & 442.5553 &  0.4967 &   3.0065 &  0.1703 & 57.7927 &     0.9085 &      1.2654 &    0.6120 &     0.6942 &    6.7434 \\[-2ex]\\
$1-\rho$ &     Cond. Std. & 9.8262 & 19.7654 & -0.0613 & 10.0610 & 442.0431 &  0.4971 &   3.0068 &  0.1708 & 57.5152 &     0.9067 &      1.2628 &    0.6121 &     0.6943 &    6.6768 \\[-2ex]\\
$1-\rho$ & Cond. Variance & 9.8406 & 19.8118 & -0.0612 & 10.0562 & 442.6330 &  0.4967 &   3.0018 &  0.1703 & 57.7828 &     0.9093 &      1.2663 &    0.6122 &     0.6944 &    6.7424 \\[-2ex]\\
$1-\rho$ &      VaR(90\%) & 9.8290 & 19.7489 & -0.0607 & 10.0559 & 442.7365 &  0.4977 &   3.0053 &  0.1712 & 57.4134 &     0.9075 &      1.2638 &    0.6122 &     0.6944 &    6.6446 \\[-2ex]\\
$1-\rho$ &      VaR(95\%) & 9.8237 & 19.7292 & -0.0608 & 10.0575 & 442.5703 &  0.4979 &   3.0098 &  0.1715 & 57.2924 &     0.9063 &      1.2621 &    0.6122 &     0.6944 &    6.6197 \\[-2ex]\\
$1-\rho$ &      VaR(99\%) & 9.8066 & 19.6849 & -0.0610 & 10.0656 & 441.5851 &  0.4982 &   3.0025 &  0.1720 & 57.0046 &     0.9055 &      1.2609 &    0.6123 &     0.6945 &    6.5649 \\[-2ex]\\
$1-\rho$ &     CVaR(90\%) & 9.8194 & 19.7209 & -0.0609 & 10.0597 & 442.2705 &  0.4979 &   3.0156 &  0.1716 & 57.2365 &     0.9056 &      1.2613 &    0.6121 &     0.6944 &    6.6068 \\[-2ex]\\
$1-\rho$ &     CVaR(95\%) & 9.8119 & 19.7012 & -0.0610 & 10.0633 & 441.8486 &  0.4980 &   2.9928 &  0.1718 & 57.1099 &     0.9070 &      1.2631 &    0.6122 &     0.6944 &    6.5839 \\[-2ex]\\
$1-\rho$ &     CVaR(99\%) & 9.7869 & 19.6531 & -0.0613 & 10.0783 & 440.0460 &  0.4980 &   2.9948 &  0.1724 & 56.7837 &     0.9055 &      1.2609 &    0.6122 &     0.6944 &    6.5810 \\[-2ex]\\
\bottomrule
\end{tabular}
}\caption*{Summary statistics for 6/6 monthly diversified reward-risk portfolios in Dow Jones Industrial Average are given in the table.  Mean and standard deviation are annualized numbers. Cumulative return is in percentage scale. Reward-risk and risk measures for 6/6 monthly diversified reward-risk portfolios in Dow Jones Industrial Average are also given. Sharpe ratio and conditional Sharpe ratio are annualized and Maximum drawdown (MDD) is in percentage scale. VaR and CVaR are represented in daily percentage scale. VaR 95/99 and CVaR 95/99 are from the ratio of 95\% to 99\%. The turnover rate is in percentage scale.}
\end{sidewaystable}

	Reward-risk profiles reported in Table \ref{tbl_summary_stats_ratio_us_dji_yf} and Table \ref{tbl_summary_stats_risk_us_dji_yf} support that the diversified reward-risk parity portfolios are less risky in various reward-risk measures than the benchmark portfolio. Most diversified reward-risk parity rules achieve higher unconditional Sharpe ratios and Calmar ratios than the equally-weighted allocation. Additionally, many alternative portfolios, in particular, all the raw allocations of the measures, also gain higher conditional Sharpe ratios. These diversified parity portfolios, except for the raw allocations of Sharpe ratio and Calmar ratio, are less exposed to downside tail risks. Besides, the portfolios constructed from the linear allocations of Calmar ratio, Sharpe ratio, STAR ratio, and risk measures exhibit not only lower maximum drawdowns but also smaller VaRs and CVaRs than the benchmark. Among the various portfolios, every risk measure of the diversified maximum drawdown and variance portfolios is at the lowest level. In particular, Sharpe ratios and Calmar ratios of these portfolios are greater than those of the benchmark, and the maximum drawdown strategies are less risky in VaR, CVaR, and maximum drawdown than the benchmark. Moreover, the inverse VaR and CVaR allocations also yield smaller downside risk measures.
	
	It is noticeable that turnover rates of the linearly-diversified reward-risk parity allocations in the Dow Jones Industrial Average universe are improved. Comparing with the benchmark turnover rate of 6.75\%, the linear diversification rules by Sharpe ratio, maximum drawdown, STAR ratio, standard deviation, variance, VaR, and CVaR are helpful to construct portfolios with reduced turnover rates of 5.49--6.69\%. In particular, the linear allocations of maximum drawdown and Sharpe ratio exhibit the lowest average turnover rates of 5.49\% and 5.79\%, respectively. It is a reminder from the previous paragraphs that the linear Sharpe ratio strategy is one of the best performers, and all the risk measures are smaller than those of the benchmark. Meanwhile, the turnover rate of the linearized Calmar ratio strategy is 26.33\%, the worst among the linear allocation strategies.
	
	However, the raw allocations of reward-risk ratio measures or the inverse allocations of pure risk measures tend to be the worst portfolios in turnover. Among these two allocation types, the raw allocation rules of reward-risk ratios tend to yield higher turnovers. The inverse diversification rules of volatility/variance and maximum drawdown yield 9.00--18.73\% and 12.29\%, respectively. In addition, turnover rates of 14.20--17.18\% are charged by the inverse allocations of VaR and CVaR. Among the raw allocations, the Rachev ratio portfolios rebalance the assets in the portfolios with relatively low turnover rates of 13.75--18.90\%. Meanwhile, the strategies from Calmar ratio and marginal/conditional Sharpe ratios replace 66.07\%, 58.46\%, and 66.84\% of notionals, respectively. The worst performers in turnover rate are all the STAR ratio portfolios of 68.58--69.20\%.
	 
\subsection{Other market universes and risk models} 
	Regardless of asset class and market, the diversified reward-risk parity strategies outperform the traditional equally-weighted allocation. In many market universes, the most profitable portfolio is constructed from the raw allocations of Calmar ratio and Sharpe ratio. However, these strategies are not the best diversification rules from the viewpoint of risk management. For example, the portfolios are heavily exposed to downside tail risks. This risky aspect of the portfolio performance is exemplified by the finding that the pure risk measures such as VaR and CVaR of the portfolios are substantially poorer than the corresponding measures of the benchmark strategy. Additionally, the turnover rates are also noticeably worse than that of the equally-weighted portfolio in each asset class. Apparently, the higher turnover rate imposes massive portfolio rebalancing that implies expensive transaction costs enough to erode considerable portions of the outstanding portfolio performance.
	
	In this perspective, using STAR ratio or Rachev ratio in portfolio construction with the raw allocation rule would be a better alternative in order to pursue the balance between performance and risk. The raw diversification rules by STAR ratio and Rachev ratio are the two most profitable portfolio construction methods that consistently outperform the equally-weighted allocation in every market covered in the paper. Moreover, higher moments such as standard deviation and kurtosis of the performance are also improved. Besides the moments of the return distributions, improved reward-risk profiles of the portfolios indicate that these portfolios are more desirable to construct not only for profitability but also for risk management. Sharpe ratios and Calmar ratios of these strategies are increased, and maximum drawdowns tend to be smaller than the benchmark drawdown. In addition, VaR and CVaR of the portfolios are slightly larger than or comparable with the risk measures of the equally-weighted allocations. Moreover, turnover rates of the diversified Rachev ratio portfolios are moderately greater in all the asset classes. Meanwhile, the worst turnover rates among all the alternative portfolios are obtained by the STAR ratio parity strategies. More importantly, the higher transaction costs caused by massive rebalancing are regarded as the trade-off for achieving the enhanced profitability and risk management.
	
	Similar to the raw allocation of Rachev ratio, the diversified reward-risk ratio parity with the linear allocation rule is the most well-balanced allocation across all the facets of portfolio management such as performance, downside risks, and transaction costs. For example, the linearly-diversified STAR ratio portfolios not only outperform the equally-weighted portfolio but also are less volatile in performance than the benchmark in all the market universes. Lower kurtosis values of the portfolios indicate that the portfolios are less exposed to heavy-tailed events than the equally-weighted portfolio. In addition to the higher moments of returns, greater Sharpe ratios of the portfolio performances also support the superiority in profitability of the linearly-diversified STAR ratio allocation rules. Moreover, the portfolios are less riskier in various risk measures than the benchmark. In various asset classes, VaR, CVaR, and maximum drawdown of the STAR ratio strategies are decreased. Lastly, turnover rates of the allocations are also reduced/comparable with respect to the equally-weighted allocation. Based on the smaller turnover rates, transaction costs can be saved even though the portfolios earn the more profitable performances under the alleviated downside risks. The similar patterns are also observed in the performance of other linearly-diversified reward-risk parity portfolios. The improved reward-risk profiles accompanied with the outperformance at the lower transaction costs are also consistent with the low volatility anomaly (\cite{Blitz:2007, Baker:2011}). 
	
	Diversified reward-risk parity using pure risk measures is also good at managing severe downside tail risks and avoiding high turnover rates. In particular, the linearly-diversified pure risk measures construct portfolios with comparable average returns and slightly lower volatility levels. For example, the portfolio performance characteristics of the VaR and CVaR allocations are almost similar to those of the linearly-diversified STAR ratio portfolios. Moreover, these strategies with enhanced Shape ratios and Calmar ratios also exhibit reduced downside risks in various risk measures. An additional advantage of these linearized risk parity allocations is taking low transaction costs. Similarly, the low-risk nature is also found in the inversely-diversified risk parity portfolios. Although the portfolios tends to be less lucrative than any other alternative portfolios, the portfolio performance yields much lower volatility levels than other portfolios including the benchmark. Additionally, Sharpe ratios and Calmar ratios of the inversely-diversified risk parity strategies are also improved, and these risk-based allocations are significantly less riskier in VaR, CVaR, and maximum drawdown than the other allocations. Meanwhile, the transaction costs are moderately worse than that of the benchmark portfolio.
	
	It is also noteworthy that the diversified reward-risk parity is robust to model selection. Through this section, we present the portfolio performance and risk profiles of the diversified reward-risk parity portfolios constructed based on reward-risk measures calculated only from the ARMA-GARCH-CTS model. However, the results is not dependent on the choice of the risk model. When the ARMA-GARCH-NTS model and the ARMA-DCC GARCH-MNTS model are used for reward-risk calculation, the similar patterns in performance and risk profiles of the diversified reward-risk parity portfolios are also found across the various market universes. Reward-risk measures from these models also construct the diversified reward-risk parity portfolios with enhanced performance, improved downside risks, and reduced transaction costs although the reward-risk profiles of those portfolios by the NTS model and the MNTS model are not presented in the paper for brevity. Additionally, the same portfolio characteristics of a given reward-risk measure under an allocation rules are observed regardless of risk model.
	
\section{Factor analysis}
\label{drrp_factor_analysis}

    In this section, we analyze the performance and risk profiles of the diversified reward-risk parity portfolios with the Carhart four-factor model. The factor analysis on the returns of the alternative portfolios in Dow Jones Industrial Average universe is conducted in order to seek better understandings on the diversified reward-risk parity portfolio construction.
    
	As seen in the previous section, the diversification portfolios of the reward-risk measures achieve the higher average returns, and are also exposed to the lower downside tail risks. The improved portfolio profitability and reward-risk profiles support the existence of the low volatility anomaly (\cite{Blitz:2007, Baker:2011}). 
	
	In order to scrutinize the low volatility anomaly observed at the diversified reward-risk allocations, the factor analysis on the portfolio returns is indispensable. Among various factor models, one of the most commonly-used factor models is the Carhart four-factor model (\cite{Carhart:1997}) that exploits four different factors: market factor (Mkt), size factor (SMB), value factor (HML), and momentum factor (MOM). The first three factors are also known as the Fama-French three factors (\cite{Fama:1993, Fama:1996}). The factor analysis on portfolio returns $r_p$ is conducted as the linear regression with respect to the four Carhart factors:
	\begin{equation}
		r_p=\alpha_p+\beta_{p,\textrm{\tiny Mkt}}f_{\textrm{\tiny Mkt}}+\beta_{p,\textrm{\tiny SMB}}f_{\textrm{\tiny SMB}}+\beta_{p,\textrm{\tiny HML}}f_{\textrm{\tiny HML}}+\beta_{p,\textrm{\tiny MOM}}f_{\textrm{\tiny MOM}}+\epsilon_p
	\end{equation}
	where $\alpha_p$ is the intercept, $\epsilon_p$ is the residual, and $\beta_{p,i}$ is the factor loading for the Carhart factor $f_i$. The historical data of the four factors were downloaded from K. R. French's data library at Dartmouth.
	
	Regression results of the Carhart four-factor analysis on the portfolio performance in the Dow Jones Industrial Average universe are given in Table \ref{tbl_carhart_factor_raw_cts_all_us_dji_yf}. It is noteworthy that many diversified reward-risk parity allocations still outperform the equally-weighted allocation even after controlling the Carhart factors. Regression intercepts of the alternative portfolios are greater than the benchmark intercept. For example, almost all the diversified risk-based parity strategies gain statistically significant and larger factor-adjusted returns than that of the benchmark. In particular, the inverse risk allocations achieve almost 6.29--33.30\%-higher alphas with statistical significance at 1\%. Intercepts of all the risk measure diversification strategies are greater than that of the benchmark. The linearly-diversified Rachev ratio portfolios mark slightly higher factor-adjusted returns than the benchmark. These larger intercepts are also consistent with the portfolio performance in section \ref{drrp_result}. Meanwhile, the linear ratio allocations by Sharpe ratio, Calmar ratio, and STAR ratio exhibit relatively smaller alphas in the factor analysis. Moreover, the Carhart four-factor intercept of the raw Sharpe ratio-based allocation is 23.53\% decreased with respect to that of the equally-weighted portfolio. Considering the outperformance of the diversified raw reward-risk ratio strategies reported in the previous section, it is obvious that large portions of the profitability earned by these raw reward-risk ratio portfolios are originated from the factor exposures.
	
\begin{table}[!ht]
\centering
\caption{Carhart 4-factor regression for CTS diversified reward-risk parity portfolios in Dow Jones Industrial Average}
\label{tbl_carhart_factor_raw_cts_all_us_dji_yf}
\resizebox{.5\paperheight}{!}{
\begin{tabular}{llllllll}
\toprule
         Rule &        Measure &    $\alpha (\%)$ &    $\beta_{Mkt}$ &    $\beta_{SMB}$ &    $\beta_{HML}$ &    $\beta_{MOM}$ &   $R^2$ \\
\midrule
            1 &   Equal Weight &   0.2129${}^{*}$ &  0.9356${}^{**}$ & -0.2271${}^{**}$ &  0.2142${}^{**}$ & -0.0754${}^{**}$ &  0.9165 \\
  $\rho|_{+}$ &         Sharpe &           0.1628 &  0.9021${}^{**}$ & -0.2200${}^{**}$ &  0.1398${}^{**}$ &  0.1253${}^{**}$ &  0.8390 \\
  $\rho|_{+}$ &   Cond. Sharpe &   0.2224${}^{*}$ &  0.9031${}^{**}$ & -0.2297${}^{**}$ &  0.1708${}^{**}$ &   0.0419${}^{*}$ &  0.8820 \\
  $\rho|_{+}$ &         Calmar &           0.2302 &  0.8923${}^{**}$ & -0.2243${}^{**}$ &  0.1284${}^{**}$ &  0.1264${}^{**}$ &  0.7900 \\
  $\rho|_{+}$ &     STAR(90\%) &   0.2277${}^{*}$ &  0.8986${}^{**}$ & -0.2327${}^{**}$ &  0.1755${}^{**}$ &   0.0468${}^{*}$ &  0.8770 \\
  $\rho|_{+}$ &     STAR(95\%) &   0.2191${}^{*}$ &  0.9004${}^{**}$ & -0.2371${}^{**}$ &  0.1705${}^{**}$ &   0.0447${}^{*}$ &  0.8743 \\
  $\rho|_{+}$ &     STAR(99\%) &   0.2091${}^{*}$ &  0.9036${}^{**}$ & -0.2394${}^{**}$ &  0.1670${}^{**}$ &   0.0478${}^{*}$ &  0.8777 \\
  $\rho|_{+}$ &   R(50\%,90\%) &  0.2136${}^{**}$ &  0.9318${}^{**}$ & -0.2304${}^{**}$ &  0.1965${}^{**}$ & -0.0486${}^{**}$ &  0.9195 \\
  $\rho|_{+}$ &   R(50\%,95\%) &   0.2089${}^{*}$ &  0.9330${}^{**}$ & -0.2328${}^{**}$ &  0.1943${}^{**}$ & -0.0500${}^{**}$ &  0.9195 \\
  $\rho|_{+}$ &   R(50\%,99\%) &   0.2007${}^{*}$ &  0.9351${}^{**}$ & -0.2336${}^{**}$ &  0.1926${}^{**}$ & -0.0480${}^{**}$ &  0.9211 \\
  $\rho|_{+}$ &   R(90\%,90\%) &  0.2166${}^{**}$ &  0.9318${}^{**}$ & -0.2288${}^{**}$ &  0.2077${}^{**}$ & -0.0614${}^{**}$ &  0.9175 \\
  $\rho|_{+}$ &   R(95\%,95\%) &  0.2160${}^{**}$ &  0.9318${}^{**}$ & -0.2297${}^{**}$ &  0.2093${}^{**}$ & -0.0660${}^{**}$ &  0.9170 \\
  $\rho|_{+}$ &   R(99\%,99\%) &  0.2196${}^{**}$ &  0.9305${}^{**}$ & -0.2297${}^{**}$ &  0.2128${}^{**}$ & -0.0734${}^{**}$ &  0.9161 \\
$1+\rho|_{+}$ &         Sharpe &   0.2126${}^{*}$ &  0.9366${}^{**}$ & -0.2266${}^{**}$ &  0.2128${}^{**}$ & -0.0696${}^{**}$ &  0.9158 \\
$1+\rho|_{+}$ &   Cond. Sharpe &   0.2119${}^{*}$ &  0.9353${}^{**}$ & -0.2281${}^{**}$ &  0.2110${}^{**}$ & -0.0673${}^{**}$ &  0.9175 \\
$1+\rho|_{+}$ &         Calmar &   0.2122${}^{*}$ &  0.9395${}^{**}$ & -0.2212${}^{**}$ &  0.1931${}^{**}$ &          -0.0127 &  0.8931 \\
$1+\rho|_{+}$ &     STAR(90\%) &   0.2129${}^{*}$ &  0.9352${}^{**}$ & -0.2278${}^{**}$ &  0.2123${}^{**}$ & -0.0691${}^{**}$ &  0.9171 \\
$1+\rho|_{+}$ &     STAR(95\%) &   0.2119${}^{*}$ &  0.9354${}^{**}$ & -0.2285${}^{**}$ &  0.2120${}^{**}$ & -0.0708${}^{**}$ &  0.9170 \\
$1+\rho|_{+}$ &     STAR(99\%) &   0.2118${}^{*}$ &  0.9356${}^{**}$ & -0.2278${}^{**}$ &  0.2129${}^{**}$ & -0.0718${}^{**}$ &  0.9170 \\
$1+\rho|_{+}$ &   R(50\%,90\%) &   0.2128${}^{*}$ &  0.9347${}^{**}$ & -0.2280${}^{**}$ &  0.2088${}^{**}$ & -0.0667${}^{**}$ &  0.9180 \\
$1+\rho|_{+}$ &   R(50\%,95\%) &   0.2111${}^{*}$ &  0.9352${}^{**}$ & -0.2287${}^{**}$ &  0.2088${}^{**}$ & -0.0683${}^{**}$ &  0.9179 \\
$1+\rho|_{+}$ &   R(50\%,99\%) &   0.2095${}^{*}$ &  0.9357${}^{**}$ & -0.2284${}^{**}$ &  0.2097${}^{**}$ & -0.0693${}^{**}$ &  0.9180 \\
$1+\rho|_{+}$ &   R(90\%,90\%) &   0.2145${}^{*}$ &  0.9339${}^{**}$ & -0.2280${}^{**}$ &  0.2109${}^{**}$ & -0.0681${}^{**}$ &  0.9173 \\
$1+\rho|_{+}$ &   R(95\%,95\%) &   0.2142${}^{*}$ &  0.9338${}^{**}$ & -0.2284${}^{**}$ &  0.2117${}^{**}$ & -0.0705${}^{**}$ &  0.9171 \\
$1+\rho|_{+}$ &   R(99\%,99\%) &   0.2163${}^{*}$ &  0.9331${}^{**}$ & -0.2285${}^{**}$ &  0.2134${}^{**}$ & -0.0744${}^{**}$ &  0.9166 \\
     $1/\rho$ &   Max Drawdown &  0.2263${}^{**}$ &  0.8772${}^{**}$ & -0.2649${}^{**}$ &  0.1822${}^{**}$ &          -0.0076 &  0.9053 \\
     $1/\rho$ &           Std. &  0.2544${}^{**}$ &  0.8606${}^{**}$ & -0.2658${}^{**}$ &  0.1801${}^{**}$ &  -0.0414${}^{*}$ &  0.9045 \\
     $1/\rho$ &       Variance &  0.2838${}^{**}$ &  0.8102${}^{**}$ & -0.2935${}^{**}$ &  0.1681${}^{**}$ &          -0.0091 &  0.8844 \\
     $1/\rho$ &     Cond. Std. &  0.2457${}^{**}$ &  0.8597${}^{**}$ & -0.2731${}^{**}$ &  0.1775${}^{**}$ &          -0.0297 &  0.9027 \\
     $1/\rho$ & Cond. Variance &  0.2690${}^{**}$ &  0.8202${}^{**}$ & -0.3021${}^{**}$ &  0.1741${}^{**}$ &           0.0087 &  0.8836 \\
     $1/\rho$ &      VaR(90\%) &  0.2516${}^{**}$ &  0.8576${}^{**}$ & -0.2730${}^{**}$ &  0.1744${}^{**}$ &          -0.0209 &  0.9026 \\
     $1/\rho$ &      VaR(95\%) &  0.2548${}^{**}$ &  0.8575${}^{**}$ & -0.2736${}^{**}$ &  0.1747${}^{**}$ &          -0.0234 &  0.9017 \\
     $1/\rho$ &      VaR(99\%) &  0.2471${}^{**}$ &  0.8602${}^{**}$ & -0.2756${}^{**}$ &  0.1739${}^{**}$ &          -0.0246 &  0.9026 \\
     $1/\rho$ &     CVaR(90\%) &  0.2502${}^{**}$ &  0.8588${}^{**}$ & -0.2743${}^{**}$ &  0.1741${}^{**}$ &          -0.0231 &  0.9025 \\
     $1/\rho$ &     CVaR(95\%) &  0.2474${}^{**}$ &  0.8597${}^{**}$ & -0.2758${}^{**}$ &  0.1735${}^{**}$ &          -0.0243 &  0.9024 \\
     $1/\rho$ &     CVaR(99\%) &  0.2425${}^{**}$ &  0.8617${}^{**}$ & -0.2764${}^{**}$ &  0.1736${}^{**}$ &          -0.0230 &  0.9029 \\
     $1-\rho$ &   Max Drawdown &   0.1992${}^{*}$ &  0.9093${}^{**}$ & -0.2480${}^{**}$ &  0.1907${}^{**}$ &  -0.0342${}^{*}$ &  0.9141 \\
     $1-\rho$ &           Std. &   0.2133${}^{*}$ &  0.9335${}^{**}$ & -0.2281${}^{**}$ &  0.2127${}^{**}$ & -0.0738${}^{**}$ &  0.9165 \\
     $1-\rho$ &       Variance &   0.2129${}^{*}$ &  0.9355${}^{**}$ & -0.2271${}^{**}$ &  0.2140${}^{**}$ & -0.0752${}^{**}$ &  0.9165 \\
     $1-\rho$ &     Cond. Std. &   0.2134${}^{*}$ &  0.9334${}^{**}$ & -0.2283${}^{**}$ &  0.2126${}^{**}$ & -0.0736${}^{**}$ &  0.9166 \\
     $1-\rho$ & Cond. Variance &   0.2129${}^{*}$ &  0.9354${}^{**}$ & -0.2271${}^{**}$ &  0.2140${}^{**}$ & -0.0753${}^{**}$ &  0.9165 \\
     $1-\rho$ &      VaR(90\%) &   0.2142${}^{*}$ &  0.9328${}^{**}$ & -0.2286${}^{**}$ &  0.2121${}^{**}$ & -0.0731${}^{**}$ &  0.9167 \\
     $1-\rho$ &      VaR(95\%) &   0.2145${}^{*}$ &  0.9319${}^{**}$ & -0.2290${}^{**}$ &  0.2114${}^{**}$ & -0.0724${}^{**}$ &  0.9167 \\
     $1-\rho$ &      VaR(99\%) &   0.2146${}^{*}$ &  0.9300${}^{**}$ & -0.2300${}^{**}$ &  0.2097${}^{**}$ & -0.0707${}^{**}$ &  0.9168 \\
     $1-\rho$ &     CVaR(90\%) &   0.2144${}^{*}$ &  0.9316${}^{**}$ & -0.2292${}^{**}$ &  0.2111${}^{**}$ & -0.0720${}^{**}$ &  0.9168 \\
     $1-\rho$ &     CVaR(95\%) &   0.2145${}^{*}$ &  0.9307${}^{**}$ & -0.2296${}^{**}$ &  0.2104${}^{**}$ & -0.0713${}^{**}$ &  0.9168 \\
     $1-\rho$ &     CVaR(99\%) &   0.2140${}^{*}$ &  0.9286${}^{**}$ & -0.2306${}^{**}$ &  0.2084${}^{**}$ & -0.0692${}^{**}$ &  0.9170 \\
\bottomrule
\end{tabular}
}\caption*{The Carhart four-factor analysis on the alternative portfolios in Dow Jones Industrial Average is given in the table.  An intercept is in monthly scale.}
\end{table}

	The patterns of the intercepts described in the above paragraph are related to the factor exposures. Although the alternative allocations and the benchmark allocation exhibit positive (negative) exposures on the market (size) and value (momentum) factors, each group of portfolios are exposed to the factors in different ways. The inverse risk strategies, which are the most profitable performers in factor-adjusted return, are about 5--20\% less exposed to the market factor and the value factor, about 50\% less exposed to the momentum factor, but about 20--40\% more exposed to the size factor than the other portfolios including the benchmark. While the market exposures of the raw Rachev ratio allocations are comparable, the exposures to the value factor and the momentum factor are 10-15\% decreased. The factor exposures of the linearized reward-risk allocations are similar to the factor structure of the equally-weighted allocation. Moreover, these factor loadings are statistically significant at 1\%, and magnitude differences in factor loading between the benchmark and the alternative allocations are relatively small. Furthermore, high $R^2$ values in the regression results of these portfolios indicate that the portfolio performances are well-explained by the Carhart factor model.
			
\section{Conclusion}
\label{drrp_conclusion}
	In this paper, we introduce alternative reward-risk parity strategies based on the diversification by reward-risk measures. The diversified reward-risk parity is the two-folded generalization of applying reward-risk measures and more generic allocation rules to traditional diversified risk parity portfolios such as the DRP portfolio and the ERC portfolio. The equally-weighted diversification is also a special case of the diversified reward-risk parity portfolio. In order to capture autocorrelation, volatility clustering, asymmetry, and heavy tails of financial time series, the reward-risk measures are calculated from three different risk models: the ARMA(1,1)-GARCH(1,1) model with CTS/NTS innovations and the ARMA(1,1)-DCC GARCH(1,1) model with MNTS innovations.
	
	Regardless of financial markets, the new heuristic reward-risk parity portfolios outperform the equally-weighted portfolio. The diversified reward-risk parity allocations are more profitable in annualized return than the benchmark. In particular, the diversification rules by Calmar ratio and Sharpe ratio are the best ways of allocating capital to increase portfolio profits. Additionally, the reward-risk portfolios constructed from Rachev ratio and STAR ratio are also improved in profitability with respect to the benchmark. Other diversified risk allocations by volatility/variance, VaR, CVaR, and maximum drawdown exhibit comparable average returns and volatility measures.
	
	The outperformance of the diversified reward-risk parity strategies is achieved by taking less downside risks. Many alternative portfolios in diverse asset classes are consistently less exposed to various risks represented by standard deviation, VaR, CVaR, and maximum drawdown than any other portfolios including the equally-weighted portfolio. Moreover, the skewness and kurtosis of the portfolio performances are improved. The Sharpe ratios and Calmar ratios of the alternative portfolios are also greater than those of the other portfolios. 
	
	In addition, the diversified reward-risk parity is robust to risk model choice. When the ARMA(1,1)-GARCH(1,1) model with NTS innovations and the ARMA(1,1)-DCC GARCH(1,1) model with MNTS innovations are used for calculating reward-risk measures in order to construct the diversified reward-risk parity portfolios, we also observe the same patterns in portfolio performance and risk profiles with the portfolios from the CTS reward-risk measures. 
	
	The Carhart four-factor analysis also supports the outperformance of the diversified reward-risk parity portfolios. After controlling the Carhart factors, the portfolio performances still remain more profitable than the equally-weighted portfolio. With high $R^2$ values, the larger regression intercepts are obtained by the diversified reward-risk allocations. The inverse risk parity portfolios achieve the largest factor-adjusted returns among the alternative portfolios, and the raw Rachev ratio portfolios also mark stronger alphas than the benchmark. Additionally, the intercepts of the linearly-diversified reward-risk portfolios consistently exceed that of the benchmark. 
	
	 In conclusion, the portfolio construction based on diversified reward-risk parity not only outperforms the equally-weighted portfolio but also provides enhanced risk management. The alternative portfolios are evidence for the low volatility anomaly.

\section*{Acknowledgement}
	The formulation of the key idea and its implementation were completed in 2014--2015 while Jaehyung Choi was attending the State University of New York at Stony Brook. The initial draft based on Bloomberg data sets by 2012 was also written at that time. Major results were recently reproduced with up-to-date data sets, and the contents of the initial draft were modified with minor revisions in accordance with the new data sets.

\end{document}